\documentclass[12pt,preprint]{aastex}
\shorttitle{Runaway Stars HD 14633 and HD 15137}
\shortauthors{Boyajian et al.}

\begin{document}

%%\received{}
%%\accepted{}

\title{The Massive Runaway Stars HD 14633 and 
HD 15137\footnote{Based in part on observations made at Observatoire de
Haute Provence (CNRS), France.}}

\author{T. S. Boyajian, T. D. Beaulieu, D. R. Gies\altaffilmark{2}, 
E. Grundstrom\altaffilmark{2}, W. Huang\altaffilmark{2},\\
M. V. McSwain\altaffilmark{2,3,4}, 
R. L. Riddle\altaffilmark{2,5}, D. W. Wingert\altaffilmark{2}}
\affil{Center for High Angular Resolution Astronomy and \\
 Department of Physics and Astronomy,\\
 Georgia State University, P. O. Box 4106, Atlanta, GA 30302-4106; \\
 boyajian@chara.gsu.edu, beaulieu@chara.gsu.edu, gies@chara.gsu.edu, 
 erika@chara.gsu.edu, huang@chara.gsu.edu, mcswain@astro.yale.edu, 
 riddle@astro.caltech.edu, wingert@chara.gsu.edu}

\altaffiltext{2}{Visiting Astronomer, Kitt Peak National Observatory,
National Optical Astronomy Observatories, operated by the Association
of Universities for Research in Astronomy, Inc., under contract with
the National Science Foundation.}

\altaffiltext{3}{Current Address: Astronomy Department,
Yale University, New Haven, CT 06520-8101}

\altaffiltext{4}{NSF Astronomy and Astrophysics Postdoctoral Fellow}

\altaffiltext{5}{Current Address: 
California Institute of Technology - TMT Project,
1200 East California Blvd., Mail Code 102-8, Pasadena, CA 91125}

\and

\author{M. De Becker} 
\affil{Institut d'Astrophysique et de G\'{e}ophysique, Universit\'{e} de Li\`{e}ge, 
17, All\'{e}e du 6 Ao\^{u}t, B5c, 4000 Sart Tilman, Belgium; \\
debecker@astro.ulg.ac.be} 

%%\slugcomment{}
%%\paperid{}

%%%%%%%%%%%%%%%%%%%%%%%%%%%%%%%%%%%%%%%%%%%%%%%%%%%%%%%%%%%%%%%

\begin{abstract}

We present results from a radial velocity study of two runaway 
O-type stars, HD~14633 (ON8.5~V) and HD~15137 (O9.5~III(n)). 
We find that HD~14633 is a single-lined spectroscopic binary 
with an orbital period of 15.4083 days.   The second target 
HD~15137 is a radial velocity variable and a possible 
single-lined spectroscopic binary with a period close to 1 month. 
Both binaries have large eccentricity, small semiamplitude, 
and a small mass function.  We show the trajectories of the 
stars in the sky based upon an integration of motion in the 
Galactic potential, and we suggest that both stars were ejected 
from the vicinity of the open cluster NGC~654 in the Perseus 
spiral arm.  The binary orbital parameters and runaway
velocities are consistent with the idea that both these stars 
were ejected by supernova explosions in binaries and
that they host neutron star companions.  We find that the 
time-of-flight since ejection is longer than the predicted 
evolutionary timescales for the stars, which may indicate 
that they have a lower mass than normally associated with 
their spectral classifications and/or that their lives 
have been extended through rapid rotation. 
 
\end{abstract}


%%%%%%%%%%%%%%%%%%%%%%%%%%%%%%%%%%%%%%%%%%%%%%%%%%%%%%%%%%%%%%%

\section{Introduction}                              % Section 1

There are two competing theories to explain the origin of 
the massive OB-runaway stars.  The model first suggested by 
\citet{zwi57} and \citet{bla61} proposes that these stars were 
originally the binary companions of a star that exploded in 
a supernova and that the linear momentum of a runaway star balances 
the momentum lost in the explosion.  Since mass ratio reversal 
probably occurs prior to the explosion, many runaways should still 
be binaries with a neutron star or black hole companion, 
unless the system was disrupted by an asymmetric kick velocity 
imparted to the remnant during the supernova \citep{bp95}. 
A second model proposes that close gravitational encounters during 
the young, high stellar number density epoch after cluster formation
can lead to ejections through encounters with hard binaries \citep{len88}.
This model predicts that most runaways will be single stars, 
although some close binaries can be ejected in exceptional 
circumstances.  \citet{gie86} made a radial velocity survey 
of bright, northern sky runaway stars and found that most were
indeed radial velocity constant, implying that they were not 
members of binary systems.   More recently \citet*{hoo00} 
explored the motions and origins of runaways using proper motion
data from the {\it Hipparcos Satellite} \citep{per97}, and   
they found examples of ejection by both mechanisms.   

Here we present new radial velocity measurements for two 
northern sky runaway stars, HD~14633 and HD~15137. 
HD~14633 is classified as a nitrogen strong ON~8V star \citep{wal72}.  
It appears at Galactic coordinates $l=140\fdg78$ and $b=-18\fdg20$, 
indicating a position nearly 800~pc below the the Galactic plane. 
The spectral lines have a moderate projected rotational velocity with
$V\sin i$ estimates of 
111 km~s$^{-1}$ \citep{con77a},
110 km~s$^{-1}$ \citep{sch88}, and 
134 km~s$^{-1}$ \citep{how97}.  \citet{rog74} found that the
star was a single-lined spectroscopic binary with a period 
of 15.335 days and an orbital eccentricity of $e=0.68$. 
However, subsequent analysis by \citet{bol78} did not 
confirm the initial orbital parameters, and \citet{bol78} 
suggested that the binary might have a nearby third star 
that modulates the velocity curve.  There is no known 
visual companion to HD~14633 \citep{mas98}.  Additional  
spectroscopic observations by \citet{sto82} showed little 
evidence of velocity variability.   

The second target is the star HD~15137 that \citet{gie87} categorized as 
a field O-star, but we show below (\S4) that its peculiar velocity 
is large enough that the star should also be grouped with the runaway stars. 
It appears in a similar part of the sky as HD~14633 at $l=137\fdg46$ and $b=-7\fdg58$.  
\citet{wal73} classified HD~15137 as O9.5~II-III~(n), where the suffix (n)
indicates broad lines.  \citet{con77a} reported observing partially resolved 
double lines in their spectrum.  However, \citet{how97} analyzed a single 
high dispersion spectrum from the {\it International Ultraviolet Explorer 
Satellite (IUE)} and used a cross-correlation method to find the 
projected rotational velocity, $V\sin i = 336$ km~s$^{-1}$. 
They caution that the cross-correlation function is broad, 
asymmetric, and difficult to measure.   We show below that the 
star is indeed broad-lined, and it may display rapid line profile 
variability normally associated with nonradial pulsation \citep{how93,kam97}. 
\citet{con77} suggest that the stellar radial velocity is variable. 
There is no evidence of a nascent cluster nearby \citep{dew04}. 

Here we present new radial velocities (\S2) based upon high S/N 
CCD spectroscopy of these two runaways.  We give new orbital 
elements for HD~14633 (\S3.1) and a tentative binary interpretation
for HD~15137 (\S3.2).  We use radial velocity and proper 
motion data to calculate the Galactic trajectories 
of both stars, and we suggest that both originated in or near the 
open cluster NGC~654 (\S4).  We argue that 
both runaways were probably ejected by a supernova in a binary
and that their unseen companions are probably neutron stars (\S5). 

%%%%%%%%%%%%%%%%%%%%%%%%%%%%%%%%%%%%%%%%%%%%%%%%%%%%%%%%%%%%%%%

\section{Observations and Radial Velocities}        % Section 2

Most of the optical spectra were obtained with the Kitt Peak National
Observatory 0.9~m Coude Feed Telescope during observing runs from 
2000 September 30 to 2000 October 13 and from
2000 December 10 to 2000 December 23. 
The spectra have a resolving power of $R=\lambda / \delta \lambda = 9500$. 
They were made using the long collimator, grating B 
(in second order with order sorting filter OG550), camera 5, and
the F3KB CCD, a Ford Aerospace $3072\times 1024$ device.
This arrangement produced a spectral coverage of $6440 - 7105$ \AA .
Exposure times varied between 20 and 30 minutes, and generally 
two spectra were taken only a few hours apart each night. 
The spectra generally have a signal-to-noise ratio of S/N $\approx 200$ pixel$^{-1}$. 
We also observed the rapidly rotating A-type star,
$\zeta$~Aql, which we used for removal of atmospheric water vapor and O$_2$ bands.
Each set of observations was accompanied by numerous bias, flat field,
and Th-Ar comparison lamp calibration frames.
One earlier red spectrum of HD~14633 was made with the Coude Feed 
Telescope on 1999 November 13, but this spectrum was obtained with
the short collimator and grating RC181 (in first order with a GG495
filter to block higher orders), which yielded a 
lower resolving power, $R=\lambda / \delta \lambda = 4000$. 
Two additional red spectra of HD~14633 were obtained with the Coude Feed 
on 2004 October 12 and 14, and one final red spectrum of 
HD~15137 was made on 2004 October 12.  
These three spectra are similar to the main group, 
but they were made with the T2KB CCD ($2048\times 2048$ pixels). 
The dates of observation are given in Tables~1 and 2.
The spectra were extracted and calibrated
using standard routines in IRAF\footnote{IRAF is distributed by the
National Optical Astronomy Observatory, which is operated by
the Association of Universities for Research in Astronomy, Inc.,
under cooperative agreement with the National Science Foundation.}.
All the spectra were rectified to a unit continuum by fitting
line-free regions.  The removal of atmospheric lines was done by
creating a library of $\zeta$~Aql spectra from each run, removing
the broad stellar features from these, and then dividing each target
spectrum by the modified atmospheric spectrum that most closely
matched the target spectrum in a selected region dominated by
atmospheric absorptions.  The spectra from each run were then
transformed to a common heliocentric wavelength grid.

\placetable{tab1}      % Table 1 - Radial Velocities for HD~14633

\placetable{tab2}      % Table 2 - Radial Velocities for HD~15137
 
Two spectra of HD~14633 in the blue domain (4550 -- 4900 \AA ) were obtained at
the Observatoire de Haute-Provence (OHP) in 2003 October. These
observations were carried out with the Aur\'{e}lie spectrograph fed by the
1.52~m telescope \citep{gil94}.  The detector was a
2048$\times$1024 CCD (EEV 42-20\#3), with a pixel size of $13.5\times13.5$ $\mu$m.
We used a 600 lines\,mm$^{-1}$ grating, offering a resolving
power of about 8000 in the blue with a reciprocal dispersion of
16 \AA ~mm$^{-1}$.  The exposure times were 45 and 30 minutes, and
the spectra have a S/N = 350 and 480 pixel$^{-1}$. 
The spectra were wavelength-calibrated using a Th-Ar
spectrum taken just after the observation of the star. 
The data were reduced using the {\sc midas} software
package developed at ESO and were normalized to a unit continuum.

We measured radial velocities for the red spectra of both HD~14633 
and HD~15137 by cross-correlating the line profiles of each spectrum 
with those in one spectrum selected for optimum S/N properties. 
We measured individually the deepest and best defined absorption lines in 
this spectral region: H$\alpha$, the blend of \ion{He}{1} $\lambda6678$ and 
\ion{He}{2} $\lambda6683$, and \ion{He}{1} $\lambda7065$.  
There was no evidence of H$\alpha$ emission in either star's spectrum.
We then formed the mean difference between the velocity for each line
and that of \ion{He}{1} $\lambda7065$, and we applied these differences
to each line's velocities to place them on the same velocity system 
as that for \ion{He}{1} $\lambda7065$ in the reference spectrum.  
Finally, we made a Gaussian fit of the \ion{He}{1} $\lambda7065$ profile
in the reference spectrum and added this to the mean velocity from 
all three lines to transform the results to absolute radial velocity.
These two stars were observed in conjunction with a program on 
eight other O-star targets, and we used measurements of the interstellar 
lines in those spectra to make small corrections (on the order of 
1 km~s$^{-1}$) to the velocity measurements from each night.  

We determined radial velocities for the two blue spectra of HD~14633 by 
parabolic fitting of the line cores of \ion{He}{1} $\lambda\lambda 4471, 4713$
and \ion{He}{2} $\lambda\lambda 4541, 4686$.  Many O-stars exhibit 
line-to-line radial velocity differences due to subtle blends and 
atmospheric expansion \citep{hut76,boh78,gie86}, but we found that the 
average radial velocity for these He lines matched those based on the 
red \ion{He}{1} $\lambda7065$ line quite well (\S3.1).  
Our final radial velocities are presented in Table 1 (HD~14633) 
and Table~2 (HD~15137).

%%%%%%%%%%%%%%%%%%%%%%%%%%%%%%%%%%%%%%%%%%%%%%%%%%%%%%%%%%%%%%%%%%%%%%%

\section{Orbital Elements}        % Section 3

\subsection{HD~14633}             % Section 3.1

\citet{rog74} found that HD~14633 is a single-lined spectroscopic binary with a 
period of 15.335~days, a small semiamplitude ($K=31.3$ km~s$^{-1}$), 
and a large eccentricity ($e=0.68$).
However, additional spectroscopic analysis by \citet{bol78} cast some doubt on the 
original solution.  Our 2000 December run was long enough to cover
an almost complete cycle of variations, and the velocities do indeed suggest an 
orbital period close to the 15~day period found by \citet{rog74}. 

We made an initial period search using the ``phase dispersion minimization'' technique
of \citet{ste78} that is especially useful for finding non-sinusoidal signals in 
time series data.  We combined our radial velocities (Table~1) with measurements 
from \citet{bol78} and \citet{sto82} (for a total of 89 measurements spanning
nearly 83 years).  We omitted from this sample three velocities from  
{\it IUE} \citep{sti01} and 
two velocities from \citet{con77} that appeared to be systematically shifted 
to more positive and more negative velocities, respectively, compared to the rest. 
We found one strong signal at a period of 15.409 days (with one weaker alias
at a period of 15.433 days), and we used this period as the starting value in 
the non-linear least squares fitting program of \citet{mor74} to establish 
the orbital elements of HD~14633.  The results are presented in Table~3 together
with the original estimates from \citet{rog74}.  The two sets of elements 
are comparable, but the new period is larger and the semiamplitude smaller 
than that obtained by \citet{rog74}.  We suspect that Rogers found an alias
period that failed to fit the additional data reported later by \citet{bol78}.  

\placetable{tab3}      % Table 3 - Orbital elements for HD 14633

The full sample of historical and new radial velocity data forms a very 
heterogeneous collection based on different lines, spectroscopic dispersions, and
S/N in the spectra.  Consequently, we repeated the orbital element fitting procedure with the more 
homogeneous set of velocities from Table~1, this time fixing the period to the 
value derived from the full, many-year sample.  These elements appear in the 
final column of Table~3, and indeed the rms residuals from the fit are now much smaller
and comparable to our measurement errors.  The final fit and observed velocities
are illustrated in Figure~1.  

\begin{figure}
\begin{center}
{\includegraphics[angle=90,height=12cm]{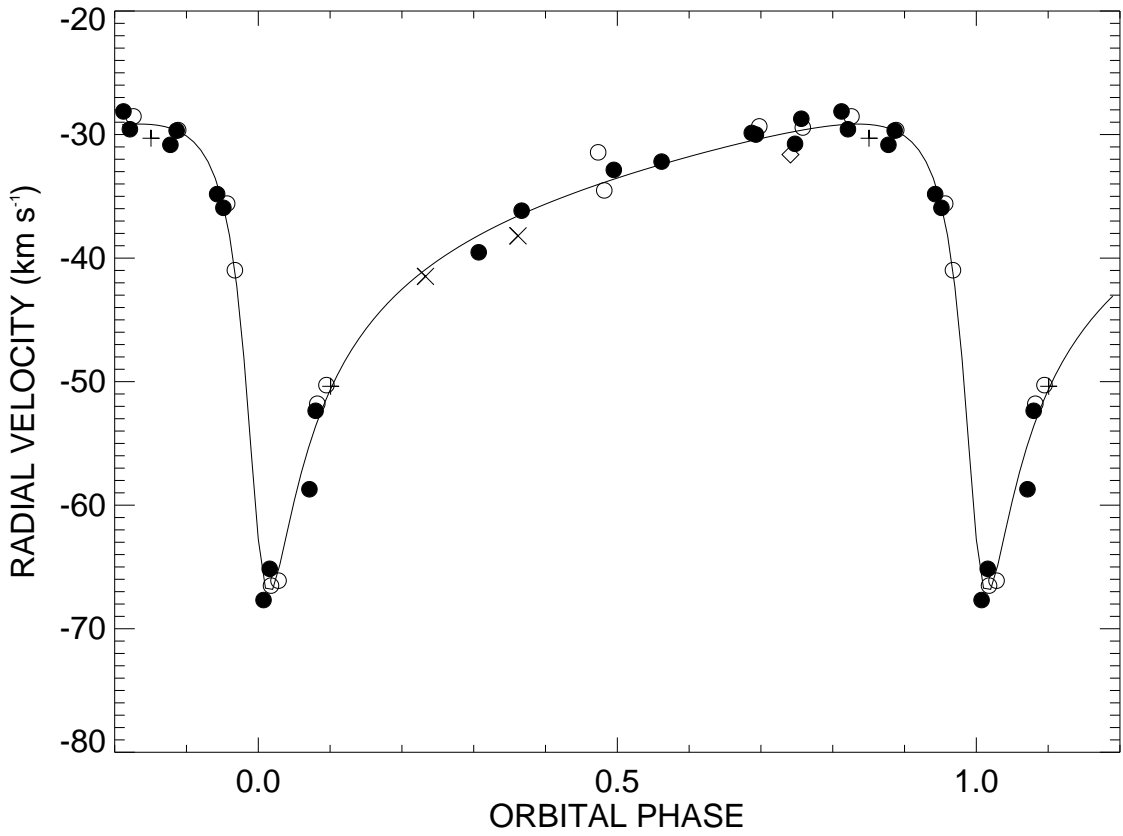}}
\end{center}
\caption{Calculated radial velocity curve ({\it solid line}) for HD 14633.
The different symbols indicate observations from 
1999 November ({\it diamond}),
2000 October ({\it open circles}),
2000 December ({\it solid circles}), 
2003 October ({\it plus signs}), and
2004 October ({\it $\times$ signs}).}
\label{fig1}
\end{figure}

\placefigure{fig1}     % Figure 1 - Radial velocity curve for HD 14633

%%%%%%%%%%%%%%%%%%%%%%%%%%%%%%%%%%%%%%%%%%%%%%%%%%%%%%%%%%%%%%%%%%%%%%%

\subsection{HD 15137}        % Section 3.2

The photospheric lines in HD~15137 are much more rotationally broadened and 
shallower than those of HD~14633.  The half-width near the continuum of 
the two \ion{He}{1} lines is $309\pm4$ km~s$^{-1}$, which is comparable to 
the projected rotational velocity of $V\sin i = 336$ km~s$^{-1}$ 
found by \citet{how97}.  The \ion{He}{1} profiles show significant 
night-to-night variations in shape that are similar to those observed in 
the nonradial pulsators HD~93521 \citep{how93} and $\zeta$~Oph \citep{kam97}, 
which are also rapidly rotating, late O-type stars.  The profiles appear with a 
central inversion on a few occasions, giving the impression of a partially resolved, 
double-lined binary (as claimed by \citealt{con77a}).  An investigation 
with a finer time resolution would clearly be rewarding, but the rapid and 
complex changes observed in the spectra available indicate that the profile 
variations are probably due to photospheric modulations rather than the 
blending of components of a short period binary.  These variations do, 
unfortunately, introduce an additional component of scatter into our 
radial velocity measurements.   Nevertheless, there is 
a clear indication that the velocity is variable on timescales of a month or so. 
The mean velocity from the 2000 October run was $-57.1 \pm 1.9$ km~s$^{-1}$
compared with $-41.6 \pm 1.0$ km~s$^{-1}$ for the 2000 December run (where 
the errors are the standard deviation of the mean).   We again used the 
phase dispersion minimization technique to search for possible periods, 
and we found candidate periods of 21.2, 28.6, and 43.4 days (with acceptable
periods in a large range surrounding the latter two).  This target 
has unfortunately been largely ignored by observers, and the only two 
measurements made in the last forty years are single velocities from 
{\it IUE} \citep{sti01} and from \citet{con77}.  Once again, the {\it IUE}
measurement appears to be much more positive than any of the other observations, 
while the measurement from \citet{con77} is lower than any of ours. 
The best fit period for our data is 28.61 days, 
but there are numerous and almost equally good alias periods at 
intervals of $+0.62 n$ days (where $n$ is an integer) spanning the
range from 28.6 to 31.1 days in addition to the other periods 
mentioned above.  We caution that the current data set samples 
essentially only the velocity extrema at two epochs, so the 
periodic nature of the variations remains to be verified. 
Nevertheless, the velocity variations are consistent with 
those expected for a long period and small semiamplitude binary. 

The limited timespan of the available data rules out the 
determination of an accurate period, but we used the 
candidate period to find a preliminary set of orbital elements. 
These elements are presented in Table~4 and the 
radial velocity curve is illustrated in Figure~2.  Although the
period is poorly known, tests with other trial periods showed that
the resulting semi-amplitude and eccentricity were not too different 
from the values reported in Table~4.  Thus, the current set of 
velocities suggests that the star is a spectroscopic 
binary with a low semiamplitude and an eccentric orbit. 

\placetable{tab4}      % Table 4 - Orbital elements for HD 15137

\begin{figure}
\begin{center}
{\includegraphics[angle=90,height=12cm]{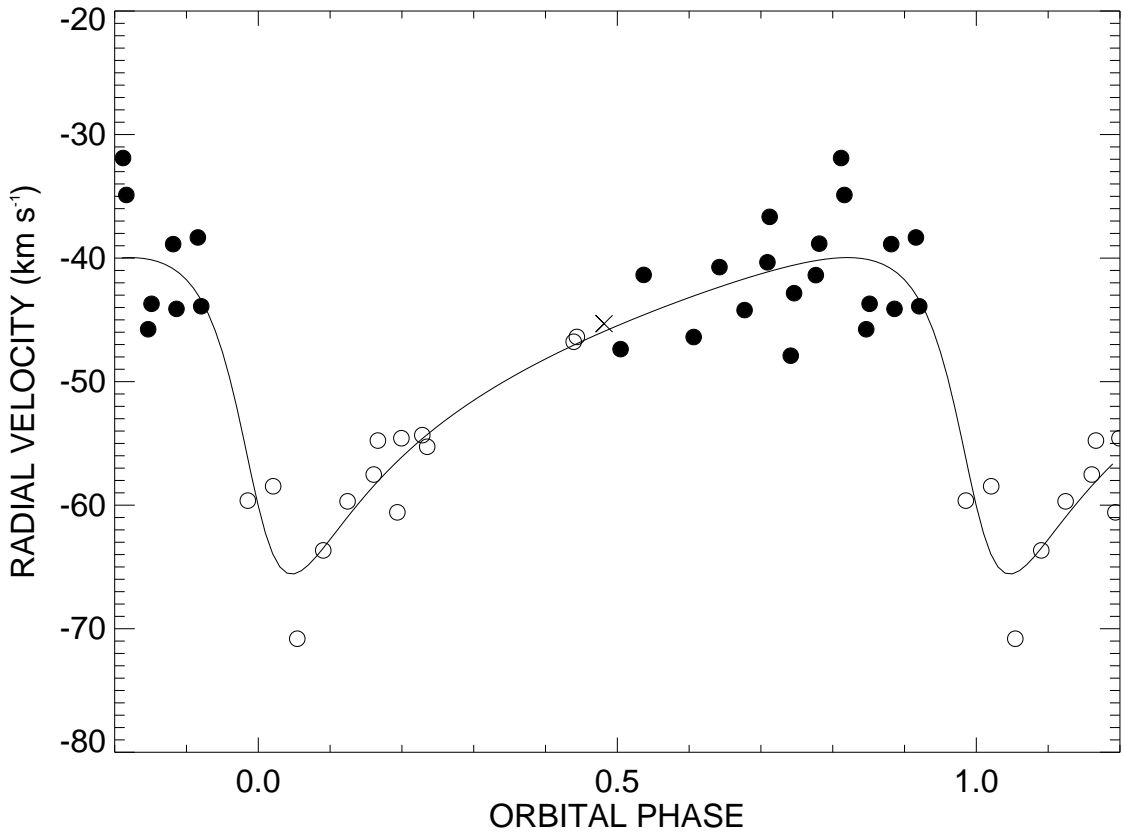}}
\end{center}
\caption{Preliminary radial velocity curve ({\it solid line}) for HD 15137.
The different symbols indicate observations from 
2000 October ({\it open circles}),  
2000 December ({\it solid circles}), and
2004 October ({\it $\times$ sign}).}
\label{fig2}
\end{figure}

\placefigure{fig2}     % Figure 2 - Radial velocity curve for HD 15137

%%%%%%%%%%%%%%%%%%%%%%%%%%%%%%%%%%%%%%%%%%%%%%%%%%%%%%%%%%%%%%%

\section{Ejection from the Galactic Plane}          % Section 4

Both HD~14633 and HD~15137 are found well outside the plane of the 
Galaxy, and {\it Hipparcos} proper motions indicate that both 
stars are moving away from the plane \citep{per97}.  Here we 
present numerical integrations of their motion in the Galaxy 
made in order to estimate their possible site of origin and their 
time-of-flight since ejection. 

The integration of motion was made using a cylindrical coordinate 
system $(r, \phi, z)$.  We first determined the position and 
resolved velocity components of the star in this system using the
Galactic coordinates $(l,b)$, proper motion, distance estimate, 
radial velocity, the velocity of the Sun with respect to the 
local standard of rest (LSR) \citep{dea98}, and the Sun's position
relative to the plane \citep*{hol97}.  We then performed integrations
backward in time using a fourth-order Runge-Kutta method and 
a model for the Galactic potential from \citet{deb98}. 
We adopted the model (\#2) from \citet{deb98} that uses a Galactocentric 
distance of 8.0~kpc and a disk density exponential scale length of 2.4~kpc. 
We used time steps of 0.01 Myr over a time span of 20 Myr.  
The procedure compared the Sun's and the star's position to find 
the distance and Galactic coordinates $l$ and $b$ for each time step. 
We determined when and where the star's trajectory crossed the Galactic plane, 
and we then integrated forward in time to find the current position and 
distance of the LSR of the intersection site.  We then inspected a list
of Galactic open clusters \citep{lei88} to search for candidate birthplace
clusters. 

We calculated a trajectory for HD~14633 using an adopted current distance 
of 2.15~kpc \citep{ste88}, the weighted mean of the 
proper motions from {\it Hipparcos} \citep{per97} and from {\it Tycho 2} \citep{hog00}, 
and the systemic radial velocity from Table~3.  According to this model, the star 
crossed the plane of the Galaxy about 13~Myr ago, in agreement with prior estimates \citep{hob83}.  
We found that the closest cluster to this trajectory was NGC~654, an open  
cluster in the Cas~OB8 association in the Perseus spiral arm. 
We calculated the trajectory of NGC~654 based on proper motions and a mean radial velocity from 
\citet*{che03} and a distance from \citet*{hue93}, and the spatial separation between 
HD~14633 and NGC~654 is plotted as a function of time in Figure~3.  
This shows that the closest approach occurred about 14.6~Myr ago.  The greatest 
uncertainty in the calculation comes from the errors in spectroscopic parallax for 
HD~14633 (approximately $\pm28\%$), so we also calculated closest separations 
for a grid of current distances to find the minimum separation possible 
with all the other parameters fixed.  We found that the minimum separation was 
11~pc for a test value of current distance of 2.24~kpc (well within the error range), 
and Figure~3 also shows the temporal variation in cluster -- star separation for this 
case.  This minimum occurred 13.9~Myr ago when the relative velocity of 
the cluster and star was 69 km~s$^{-1}$.  If the star was actually ejected 
at this time from this cluster, then this relative velocity is the ejection velocity. 

\begin{figure}
\begin{center}
{\includegraphics[angle=90,height=12cm]{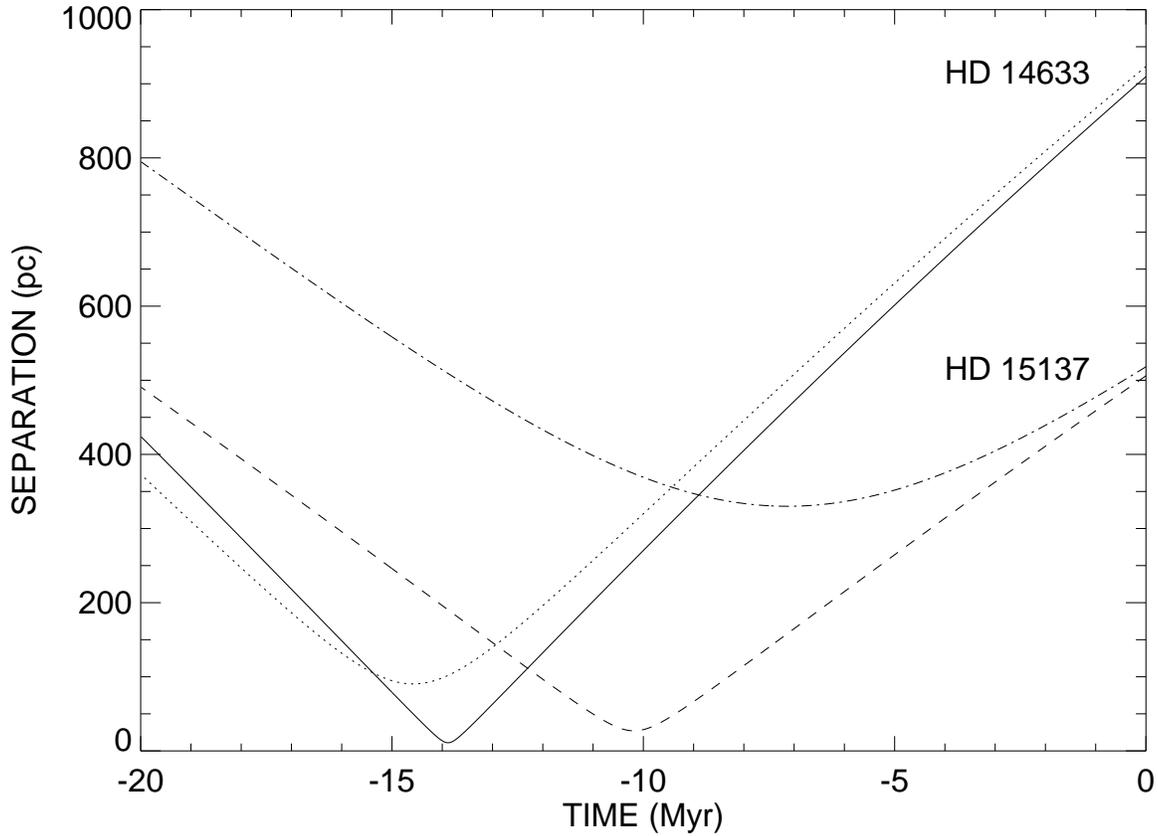}}
\end{center}
\caption{The separation between HD~14633 and NGC~654 and 
between HD~15137 and NGC~654 plotted against time 
in millions of years relative to the present.  
The dotted line shows the separation 
for the nominal current distance to HD~14633 of 2.15~kpc 
while the solid line shows the separation for a trajectory 
calculated using a current distance of 2.24~kpc. 
Likewise the dot-dashed line shows the separation for the nominal
distance to HD~15137 of 2.65~kpc while the dashed line shows the
same for a current distance of 2.29~kpc.}
\label{fig3}
\end{figure}

\placefigure{fig3}     % Figure 3 - Separation of HD 14633, HD15137 from NGC 654

We illustrate the trajectories of the star and cluster as viewed from the Sun 
in Figure~4.  Tick marks along each trajectory
mark intervals of 1~Myr before the current time ({\it diamonds}).  
We also show the trajectories for the $\pm1 \sigma$ errors in the
proper motions ({\it dotted lines}).  The errors in 
proper motion probably introduce a $\pm 2$~Myr error in the estimated time
of closest approach. 

\begin{figure}
\begin{center}
{\includegraphics[angle=90,height=12cm]{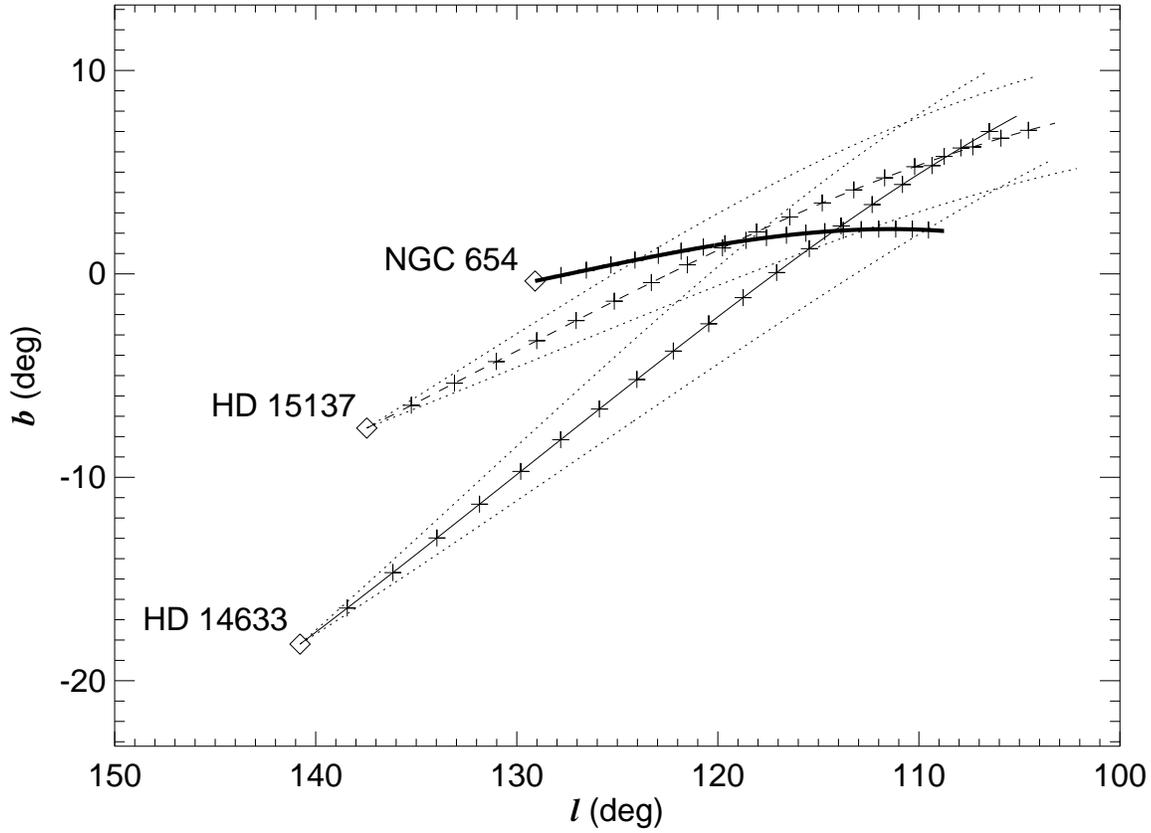}}
\end{center}
\caption{The past trajectories of HD~14633 ({\it thin solid line}), 
HD~15137 ({\it thin dashed line}), and NGC~654 ({\it thick solid line}) 
in Galactic longitude and latitude. 
The diamonds mark the current positions and tick marks are placed 
at 1~Myr intervals along each track.  The dotted lines 
show the tracks for $\pm1 \sigma$ errors in proper motion for the runaway stars.}
\label{fig4}
\end{figure}

\placefigure{fig4}     % Figure 4 - Trajectories of HD 14633, HD15137, and NGC 654

We performed the same kind of calculation for HD~15137 using a nominal 
distance estimate of 2.65~kpc \citep{ste88}, the 
weighted mean of the {\it Hipparcos} and {\it Tycho 2} proper 
motions, and the systemic radial velocity from Table~4.  We found the 
star crossed the plane of the Galaxy some 8~Myr ago for this assumed distance. 
We searched for possible clusters of origin, and we were surprised to 
find that NGC~654 once again presented the closest approach of trajectories. 
The separation between HD~15137 and NGC~654 is plotted in Figure~3, 
and we found that the smallest separation was 328~pc for the nominal distance
estimate.  However, we tested a grid of trajectories for different values of the 
assumed current distance, and the minimum star -- cluster separation 
occurred for an assumed current distance of 2.29~kpc (again within the errors
associated with the spectroscopic parallax).  The minimum separation 
was 27~pc at a time 10.2~Myr ago when the relative velocity was 50 km~s$^{-1}$
(Fig.~3).   The paths of the star and cluster for the past 20~Myr are 
illustrated in Figure~4, where we see that errors in the proper motion 
contribute an uncertainty of $\pm 2$~Myr in the estimate of the ejection time. 

%%%%%%%%%%%%%%%%%%%%%%%%%%%%%%%%%%%%%%%%%%%%%%%%%%%%%%%%%%%%%%%

\section{Discussion}                                % Section 5

OB runaway stars are probably ejected by one of two mechanisms, 
sudden mass loss during a supernova explosion in a binary or
a close gravitational encounter involving binaries \citep{gie86,hoo00}. 
The supernova theory predicts that runaways will either be 
single stars (in which the binary was disrupted due to a large, asymmetric
kick velocity imparted during the supernova) or binaries with 
neutron star or black hole companions (such as the high mass 
X-ray binaries).  On the other hand, the gravitational encounter 
theory suggests that most runaways will be single objects, 
although in rare cases hard binaries of mass ratio near unity 
are ejected.   

Our radial velocity study has demonstrated that HD~14633 and 
possibly HD~15137 are binary stars with low mass companions. 
If we suppose that the masses of the primary are $23~M_\odot$ 
for HD~14633 \citep{kee84} and $24~M_\odot$ for HD~15137
\citep*{vac96}, then the minimum masses of the companion
derived from the orbital mass function (Tables 3 and 4)  
will be $1.3~M_\odot$ and $1.5~M_\odot$, respectively 
(for an orbital inclination of $90^\circ$).  
These masses are close to the $1.35~M_\odot$ value found 
for most neutron stars in binaries \citep{tho99}. 
These runaways may be the first examples of the long sought ``quiet''
massive X-ray binaries, i.e., those with wide separations in which wind accretion 
is too weak to power an accretion disk X-ray source \citep{vdh76}.   
We searched for evidence of a companion spectrum in both 
cases using a Doppler tomography algorithm \citep{bag94},
but no spectral features were found.  A faint, low mass, main 
sequence star could easily remain hidden in the glare of an O-star 
(for example, the magnitude difference is $\triangle V \approx 8$ 
between such O-stars and a F3~V companion of mass $1.4~M_\odot$).
Nevertheless, we doubt that these systems are extreme 
mass ratio binaries containing an O- and F-type star, since
no such systems are known among the O-stars and since such systems 
would probably be disrupted in close gravitational encounters 
leading to ejection.  

Both HD~14633 and HD~15137 have many characteristics in common
with the massive X-ray binary and microquasar, LS~5039 \citep{mcs04}.
All are runaway objects with very eccentric orbits and small 
orbital mass functions.   LS~5039 has a much shorter period 
(4.4267 days) and the smaller semimajor axis results in a 
modestly dense wind in the vicinity of the orbiting neutron star, 
so that LS~5039 is a weak X-ray source.  In contrast, 
the longer period systems HD~14633 and HD~15137
will have very rarefied winds close to their neutron star 
companions, and consequently their accretion fluxes are 
expected to be extremely faint (perhaps also as the result
of centrifugal inhibition of accretion; \citealt*{ste86}). 
Neither system is listed in the ROSAT All-Sky Survey 
Faint Source Catalogue \citep{vog00}.  Furthermore, neither 
system appears to be associated with an {\it EGRET} 
$\gamma$-ray source \citep{har99}, nor are they known radio 
sources \citep*{val85,wen95,say96}.  Thus, wind accretion onto a neutron star 
in these systems must be too feeble to produce the high energy 
phenomena associated with other massive X-ray binaries. 

\citet{mcs04} found that 
the supernova mass loss prediction for LS~5039 was different 
depending on whether the calculation was based on orbital eccentricity or
runaway velocity, and they argued that both the eccentricity and 
runaway velocity can be explained if a significant asymmetric kick 
velocity was imparted to the neutron star during formation.  
A similar conclusion can be derived for HD~14633 and HD~15137.
If we use the expressions for supernova mass loss given by \citet*{nel99} 
and adopt the primary masses given above and secondary masses of $1.4~M_\odot$,
then the predicted supernova mass loss is $17~M_\odot$ and $13~M_\odot$ 
for HD~14633 and HD~15137, respectively, based upon their observed eccentricities.
On the other hand, the supernova mass loss estimates  
are $6.9~M_\odot$ and $6.4~M_\odot$, respectively, 
based upon the relative runaway velocities between star and cluster 
from the models given in \S4.  These significant differences suggest 
that both systems suffered kick velocities at birth that substantially 
altered the eccentricity.  The supernova mass loss estimates from the 
runaway velocities should be more reliable since the runaway velocities
are less affected by kicks \citep{bp95}. 

Two other features of these stars also link them to supernova ejections. 
First, HD~14633 is a well known nitrogen rich ON star \citep{wal72,sch88}, 
and \citet{mcs04} have shown that massive X-ray binary LS~5039 also 
shares this trait.  \citet{mcs04} suggest that the nitrogen enrichment 
is the result of mass transfer of CNO-processed gas from the supernova 
progenitor prior to the explosion, although rotationally induced mixing
may also play a role.   Second, HD~15137 is a very rapid rotator, 
a characteristic shared with many other OB runaway stars \citep{bla93}. 
Mass transfer prior to the supernova may lead to a spin up of the mass gainer, 
and this process may be responsible for the largest class of massive
X-ray sources, the rapidly rotating, Be X-ray binaries \citep{coe00}.    

Both runaways appear to have been ejected from the Perseus spiral arm, 
and our analysis of their motions in the Galaxy (\S4) indicates a 
probable origin in the open cluster NGC~654 in the Cas~OB8 association. 
The cluster's age is approximately 14~Myr \citep{hue93}, and the cluster contains 
a number of early B-type stars and two massive supergiants (HD~10494, F5~Ia, and 
BD$+61^\circ315$, A2~Ib).  \citet{gar92} include the nearby O-star  
BD$+60^\circ261$ (O7.5~III(n)((f)); \citealt{wal73}) as a cluster member. 
The Cas~OB8 association has a diameter of approximately 85~pc and contains 
several other clusters including NGC~581, 659, and 663 \citep{gar92},
which have slightly greater ages of 22, 35, and 16~Myr, respectively, 
according to the Webda database\footnote{Maintained by J.-C. Mermilliod at
http://obswww.unige.ch/webda/webda.html} \citep{mer03}. 
The time-of-flight for HD~15137 (10~Myr) suggests that the star was ejected 
from NGC~654 when the cluster was approximately 4~Myr old, which may be 
consistent with the evolutionary timescale required for a supernova progenitor. 
However, the main sequence lifetime of a star of $24~M_\odot$ is 
approximately 6.7~Myr \citep{sch92}, which is less than the time-of-flight
for HD~15137.  The situation is even more discrepant for HD~14633 which 
has a time-of-flight of at least 12~Myr (see also the extreme case of 
the runaway star HD~93521; \citealt{how93}).  It is difficult to reconcile 
these long travel times with the expected short lifetimes of O-stars.  
There are several possible explanations.  First, the runaways may 
have been rejuvenated by mass transfer just prior to the supernova 
explosion, which would reset their effective zero-age times to 
an epoch just prior to ejection.  Second, at least HD~15137 
is a rapid rotator, and fast rotation may help to mix gas and 
extend the main sequence lifetime of massive stars \citep{heg00,mey00}. 
Third, these stars may be over-luminous for their mass in same way 
as found for some massive X-ray binaries \citep{kap01}, so that their 
masses are lower and evolutionary lifetimes longer than simple estimates suggest.

The orbital properties of these two runaway binaries, their 
small mass functions, and their probable origin in a cluster 
containing evolved, massive stars all indicate that these stars 
were ejected during a supernova explosion in a binary.  
They are not known X-ray sources, due presumably to their 
large semimajor axes and low wind accretion rates, but it is 
possible that they exhibit transient X-ray emission when 
their neutron stars pass through the densest stellar wind 
regions near the periastron orbital phase.  
It is important to pursue radial velocity studies of other OB runaway 
stars to search for additional instances of such low amplitude 
binary systems.  Only then will we determine the relative importance 
of the supernova and close encounter ejection processes for the kinematics of 
massive stars.  

%%%%%%%%%%%%%%%%%%%%%%%%%%%%%%%%%%%%%%%%%%%%%%%%%%%%%%%%%%%%%%%

\acknowledgments

We thank the staff of KPNO for their assistance in making 
these observations possible.   We also thank Walter Dehnen for 
sending us his code describing the Galactic gravitational potential. 
Financial support was provided by the National Science 
Foundation through grant AST$-$0205297 (DRG).
Institutional support has been provided from the GSU College
of Arts and Sciences and from the Research Program Enhancement
fund of the Board of Regents of the University System of Georgia,
administered through the GSU Office of the Vice President
for Research.  MD acknowledges
financial support through the PRODEX XMM-OM Project.

%%%%%%%%%%%%%%%%%%%%%%%%%%%%%%%%%%%%%%%%%%%%%%%%%%%%%%%%%%%%%%%

% References

%%%%%%%%%%%%%%%%%%%%%%%%%%%%%%%%%%%%%%%%%%%%%%%%%%%%%%%%%%%%%%%
% Tables

\clearpage

% Table 1
\begin{deluxetable}{lccc}
\tabletypesize{\scriptsize}
\tablewidth{0pt}
\tablenum{1}
\tablecaption{HD 14633 Radial Velocity Measurements\label{tab1}}
\tablehead{
\colhead{HJD}             &
\colhead{Orbital}         &
\colhead{$V_r$}           &
\colhead{$O-C$}           \\
\colhead{($-$2,400,000)}  &
\colhead{Phase}           &
\colhead{(km s$^{-1}$)}   &
\colhead{(km s$^{-1}$)}   }
\startdata
  51495.903 &  0.741 &    $-$31.6 &   $-$1.8 \\
  51818.807 &  0.698 &    $-$29.3 &  \phs1.0 \\
  51819.741 &  0.758 &    $-$29.4 &  \phs0.2 \\
  51820.786 &  0.826 &    $-$28.5 &  \phs0.6 \\
  51821.746 &  0.888 &    $-$29.6 &  \phs0.1 \\
  51822.797 &  0.957 &    $-$35.6 &  \phs1.6 \\
  51822.963 &  0.967 &    $-$41.0 &  \phs0.1 \\
  51823.738 &  0.018 &    $-$66.5 &  \phs0.5 \\
  51823.899 &  0.028 &    $-$66.1 &   $-$0.9 \\
  51824.732 &  0.082 &    $-$51.8 &  \phs1.4 \\
  51824.928 &  0.095 &    $-$50.3 &  \phs1.1 \\
  51830.758 &  0.473 &    $-$31.4 &  \phs2.6 \\
  51830.889 &  0.482 &    $-$34.5 &   $-$0.6 \\
  51889.828 &  0.307 &    $-$39.5 &   $-$1.4 \\
  51890.749 &  0.367 &    $-$36.2 &  \phs0.3 \\
  51892.727 &  0.495 &    $-$32.9 &  \phs0.8 \\
  51893.753 &  0.562 &    $-$32.2 &  \phs0.2 \\
  51895.688 &  0.687 &    $-$29.9 &  \phs0.6 \\
  51895.777 &  0.693 &    $-$30.0 &  \phs0.4 \\
  51896.617 &  0.748 &    $-$30.8 &   $-$1.0 \\
  51896.750 &  0.756 &    $-$28.7 &  \phs0.9 \\
  51897.613 &  0.812 &    $-$28.1 &  \phs1.1 \\
  51897.751 &  0.821 &    $-$29.6 &   $-$0.4 \\
  51898.620 &  0.878 &    $-$30.8 &   $-$1.4 \\
  51898.754 &  0.886 &    $-$29.7 &  \phs0.0 \\
  51899.624 &  0.943 &    $-$34.8 &   $-$0.8 \\
  51899.757 &  0.951 &    $-$35.9 &   $-$0.2 \\
  51900.619 &  0.007 &    $-$67.7 &   $-$1.6 \\
  51900.751 &  0.016 &    $-$65.2 &  \phs2.0 \\
  51901.604 &  0.071 &    $-$58.7 &   $-$3.6 \\
  51901.737 &  0.080 &    $-$52.4 &  \phs1.3 \\  
  52930.558 &  0.851 &    $-$30.3 &   $-$1.1 \\ 
  52934.413 &  0.101 &    $-$50.4 &  \phs0.2 \\ 
  53290.840 &  0.233 &    $-$41.5 &   $-$0.6 \\
  53292.825 &  0.362 &    $-$38.2 &   $-$1.6 \\
\enddata
\end{deluxetable}

\newpage

%Table 2 

\begin{deluxetable}{lccc}
\tabletypesize{\scriptsize}
\tablewidth{0pt}
\tablenum{2}
\tablecaption{HD 15137 Radial Velocity Measurements\label{tab2}}
\tablehead{
\colhead{HJD}             &
\colhead{Orbital}         &
\colhead{$V_r$}           &
\colhead{$O-C$}           \\
\colhead{($-$2,400,000)}  &
\colhead{Phase}           &
\colhead{(km s$^{-1}$)}   &
\colhead{(km s$^{-1}$)}   }
\scriptsize
\startdata
  51817.788 &  0.985 &  $-$59.6 &   $-$3.4 \\
  51818.797 &  0.021 &  $-$58.5 &  \phs5.6 \\
  51819.758 &  0.054 &  $-$70.8 &   $-$5.3 \\
  51820.793 &  0.090 &  $-$63.7 &   $-$0.2 \\
  51821.762 &  0.124 &  $-$59.7 &  \phs1.3 \\
  51822.805 &  0.161 &  $-$57.5 &  \phs0.9 \\
  51822.970 &  0.167 &  $-$54.8 &  \phs3.3 \\
  51823.746 &  0.194 &  $-$60.6 &   $-$4.1 \\
  51823.907 &  0.199 &  $-$54.6 &  \phs1.6 \\
  51824.739 &  0.228 &  $-$54.3 &  \phs0.3 \\
  51824.935 &  0.235 &  $-$55.3 &   $-$1.0 \\
  51830.770 &  0.439 &  $-$46.8 &  \phs0.3 \\
  51830.897 &  0.444 &  $-$46.4 &  \phs0.5 \\
  51889.849 &  0.505 &  $-$47.4 &   $-$2.0 \\
  51890.769 &  0.537 &  $-$41.4 &  \phs3.2 \\
  51892.761 &  0.606 &  $-$46.4 &   $-$3.3 \\
  51893.788 &  0.642 &  $-$40.7 &  \phs1.6 \\
  51894.790 &  0.677 &  $-$44.2 &   $-$2.5 \\
  51895.699 &  0.709 &  $-$40.3 &  \phs0.8 \\
  51895.788 &  0.712 &  $-$36.7 &  \phs4.4 \\
  51896.628 &  0.742 &  $-$47.9 &   $-$7.3 \\
  51896.762 &  0.746 &  $-$42.8 &   $-$2.3 \\
  51897.627 &  0.776 &  $-$41.4 &   $-$1.2 \\
  51897.762 &  0.781 &  $-$38.8 &  \phs1.3 \\
  51898.633 &  0.812 &  $-$31.9 &  \phs8.1 \\
  51898.766 &  0.816 &  $-$34.9 &  \phs5.1 \\
  51899.635 &  0.847 &  $-$45.8 &   $-$5.7 \\
  51899.768 &  0.851 &  $-$43.7 &   $-$3.6 \\
  51900.630 &  0.881 &  $-$38.9 &  \phs2.0 \\
  51900.762 &  0.886 &  $-$44.1 &   $-$3.1 \\
  51901.615 &  0.916 &  $-$38.3 &  \phs4.6 \\
  51901.748 &  0.921 &  $-$43.9 &   $-$0.5 \\
  53290.849 &  0.482 &  $-$45.3 &  \phs0.6 \\
\enddata
\end{deluxetable}

\newpage

%Table 3 
\begin{deluxetable}{lccc}
%\tabletypesize{\scriptsize}
\tablewidth{0pc}
\tablenum{3}
\tablecaption{Orbital Elements for HD 14633\label{tab3}}
\tablehead{
\colhead{} & 
\colhead{} &
\colhead{Bolton \& Rogers (1978) +} &
\colhead{}\\
\colhead{Element} & 
\colhead{Rogers (1974)} &
\colhead{Stone (1982) + New} &
\colhead{New}}
\startdata
$P$~(days)              \dotfill    & 15.335  & $15.4083 \pm 0.0004$ & 15.4083\tablenotemark{a}  \\
$T$ (HJD--2,400,000)    \dotfill    & 42007.3 & $44227.26 \pm 0.21$  & $51854.28 \pm 0.05$       \\
$e$                     \dotfill    & 0.68    & $0.63 \pm 0.05$      & $0.698 \pm 0.010$         \\
$\omega$ (deg)          \dotfill    & 166.3   & $142 \pm 10$         & $140.3 \pm 2.2$           \\
$K$ (km s$^{-1}$)       \dotfill    & 31.3    & $15.9 \pm 1.4$       & $19.0 \pm 0.4$            \\
$\gamma$ (km s$^{-1}$)  \dotfill    & $-46.0$ & $-38.8 \pm 0.8$      & $-37.9 \pm 0.3$           \\
$f(m)$  ($M_\odot$)     \dotfill    & 0.019   & $0.0030 \pm 0.0009$  & $0.0040 \pm 0.0003$       \\
$a_1\sin i$ ($R_\odot$) \dotfill    & 6.95    & $3.8 \pm 0.4$        & $4.14 \pm 0.10$           \\
rms (km s$^{-1}$)       \dotfill    & \nodata & $7.4$                & $1.3$                     \\
\enddata
\tablenotetext{a}{Fixed.}
\end{deluxetable}

\newpage

%Table 4 
\begin{deluxetable}{lc}
%\tabletypesize{\scriptsize}
\tablewidth{0pc}
\tablenum{4}
\tablecaption{Preliminary Orbital Elements for HD 15137\label{tab4}}
\tablehead{
\colhead{Element} & 
\colhead{Value}}
\startdata
$P$~(days)              \dotfill    & $28.61 \pm 0.09$          \\
$T$ (HJD--2,400,000)    \dotfill    & $51904.0 \pm  0.7$        \\
$e$                     \dotfill    & $0.52 \pm 0.07$           \\
$\omega$ (deg)          \dotfill    & $125 \pm 11$              \\
$K$ (km s$^{-1}$)       \dotfill    & $12.9 \pm 1.3$            \\
$\gamma$ (km s$^{-1}$)  \dotfill    & $-49.0 \pm 0.7$           \\
$f(m)$  ($M_\odot$)     \dotfill    & $0.0039 \pm 0.0013$       \\
$a_1\sin i$ ($R_\odot$) \dotfill    & $6.2 \pm 0.7$             \\
rms (km s$^{-1}$)       \dotfill    & $3.8$                     \\
\enddata
\end{deluxetable}

\newpage

%%%%%%%%%%%%%%%%%%%%%%%%%%%%%%%%%%%%%%%%%%%%%%%%%%%%%%%%%%%%%%

% Figure captions

\clearpage

\input{epsf}
% Figure 1

% Figure 2

% Figure 3

% Figure 4

%%%%%%%%%%%%%%%%%%%%%%%%%%%%%%%%%%%%%%%%%%%%%%%%%%%%%%%%%%%%%%%

\end{document}